\documentclass[aps,amsmath,amssymb,reprint,twocolumn,floatfix]{revtex4-2}

\usepackage{graphicx}
\usepackage{bm}

\usepackage[utf8]{inputenc}
\usepackage[T1]{fontenc}
\usepackage{mathptmx}
\usepackage{etoolbox}
\usepackage{siunitx} 
\def\unit#1{{\;\mathrm{#1} }}

\def\vv{{\bf v}}
\def\kk{{\bf k}}
\def\FF{{\bf F}}

\def\khat{{\bf \hat{k}}}
\def\pp{{\bf p}}

\def\eg{{\em e.g.~}}

\def\exhat{{ \bf\hat{e}}_1}
\def\eyhat{{ \bf\hat{e}}_2}
\def\ezhat{{ \bf\hat{e}}_3}
\def\exhatf{{ \hat{e}}_1}

\def\nhat{{\bf \hat{n}}}

\def\etal{{\em et al. }}
\def\eg{{\em e.g. }}

\def\frameM{$\cal{M}$}
\def\frameL{$\cal{L}$}

\begin{document}


\title{Poynting-Robertson damping of laser beam driven lightsails }
\author{Rhys Mackintosh}
 \author{Jadon Y. Lin}
 \author{Michael S. Wheatland}
\affiliation{ 
 The University of Sydney, School of Physics, NSW 2006 Camperdown, Australia
}

 \author{Boris T. Kuhlmey}
 
\email{boris.kuhlmey@sydney.edu.au}
\affiliation{The University of Sydney, School of Physics, Institute of Photonics and Optical Science, and The University of Sydney Nano Institute,  NSW 2006 Camperdown, Australia}

\date{\today}

\begin{abstract}
Lightsails  using  Earth-based lasers  for propulsion   require passive stabilization to stay within the beam. This can be achieved through the sail's scattering properties, creating optical restoring forces and torques. Undamped restoring forces produce uncontrolled oscillations, which could jeopardize the mission, but it is not obvious how to achieve damping in the vacuum of space.
Using a simple two-dimensional model  we show that the Doppler effect and relativistic aberration of the propelling laser beam  create damping terms in the optical forces and torques. The effect is  similar  to  the Poynting-Robertson effect causing loss of orbital momentum of dust particles around stars, but can be enhanced by design of the sail's geometry.
\end{abstract}

\maketitle

\section{\label{sec:introduction}Introduction}
Laser powered lightsails~\cite{Marx1966,Forward1984} are  one of the few plausible pathways for sending probes to other stars on time scales of a single human generation. With physically realistic but extremely challenging infrastructure~\cite{Parkin2018,lubin2022roadmap}, a lightsail of mass $m\simeq 1\unit{g}$ could be accelerated to a velocity $v=0.2c$ within a time $\sim 1000\unit{s}$ and acceleration distance $\lesssim 0.1\unit{AU}$, reaching the Proxima Centauri system within  20 years. Such a lightsail would be propelled by a powerful laser array based on Earth, the photons of the laser imparting  momentum upon reflection on the sail. Because of the laser beam's  finite width, a mechanism is required for the lightsail to remain in the center of the beam. Any feedback to adjust the ground-based laser would be too slow as soon as the lightsail is a few light-milliseconds away, and active impulse or optical feedback mechanisms on board are difficult to achieve within the mass budget and without adding optical absorption that could lead to thermal breakdown of the sail. The most likely implementation of stabilization is thus through passive optical stabilization~\cite{Ilic2019,Srivastava2019,Salary2020,Kumar2021,Rafat2022}, which uses the combination of beam shape and spatial  reflectivity profile to generate a spring-like restoring force towards the  center of the beam, as well as a restoring torque to keep the sail at the optimal angle. However, the restoring force and torque alone lead to oscillations, which in the absence of damping are maintained throughout the acceleration phase, with an amplitude  likely to increase with any perturbations due to, for example, time-dependent beam misalignment during the acceleration phase. 

\begin{figure}[hbtp]
    \centering
    \includegraphics[width=0.5\textwidth]{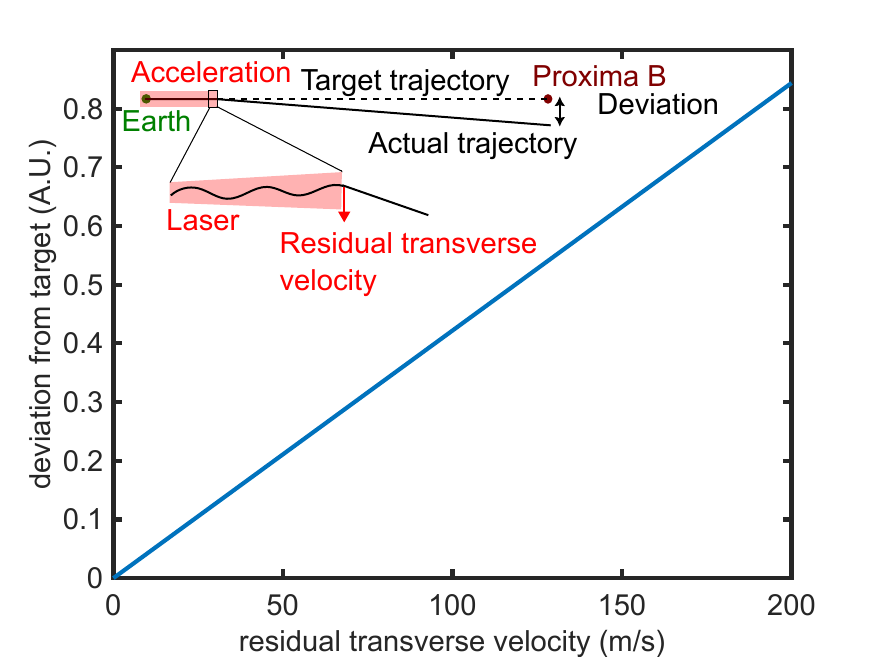}
    \caption{Deviation from target course in astronomical units after 20 years of travel as a function of residual unwanted transverse velocity. }
    \label{fig:deviation from course}
\end{figure}

Any  transverse velocity component remaining at the end of the acceleration phase will lead to the craft going off-course,  with dramatic consequences on the ability to take  telemetric measurements of exo-planets. Figure~\ref{fig:deviation from course} shows the deviation  from the ideal trajectory after 20 years of cruising at 0.2c as a function of residual  transverse velocity at the end of the acceleration phase. Simulations of passive stabilization using optical forces showed final residual transverse velocities of order $\sim 1-150\unit{m/s}$, depending on implementation and level of perturbations~\cite{Salary2020,Ilic2019,Rafat2022}, leading to final deviations of up to 0.6 astronomical units after 20 years of travel.
Equally, any residual angular velocity is likely to  complicate both telemetry and communications with Earth. 

Damping is required to reduce these oscillations, but  is difficult to achieve in space.  Srivastava \etal~\cite{Srivastava2019} included an arbitrary damping force term without justifying what physical mechanism may cause it. Salary \etal~\cite{Salary2020} saw a reduction in spatial amplitude of the oscillations in their simulations,  attributed to the shift in frequency from the Doppler effect changing the reflectivity and thus restoring force. This is akin to changing the stiffness of a spring in a mass-spring system, and thus to lowest order like changing the slope of the associated parabolic potential well: the spatial extent of the oscillation is reduced, but the total energy and thus maximum kinetic energy during the oscillations remain  unaffected. It is thus unclear how much the Doppler effect in that situation is reducing the {\em velocity} amplitude of the oscillations. Rafat \etal~\cite{Rafat2022} proposed the use of a damped internal degree of freedom, which can indeed  effectively reduce the oscillations of the lightsail, both in spatial amplitude and in velocity. However the implementation of a damped internal degree of freedom will be challenging within the mass budget of a lightsail only hundreds of nm thick.

The Doppler effect  has been used as a damping force to slow  atoms~\cite{hansch1975}, but only in the direction of propagation of the laser.
It is difficult to see how this could be implemented to dampen transverse oscillation of a lightsail, as it appears it would require laser beams  propagating orthogonal to the direction of acceleration.

\begin{figure}[htb]
        \centering
        \includegraphics{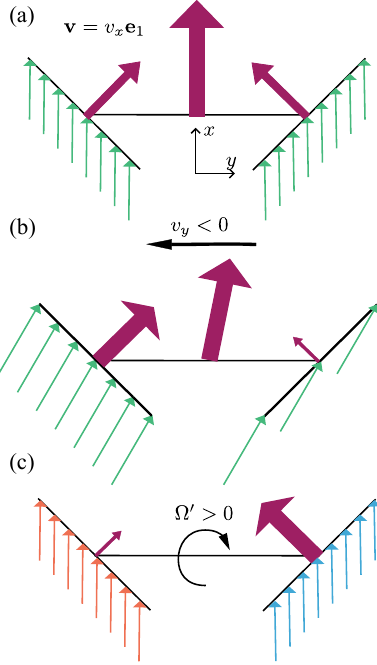}
        \caption{Principle of PR damping of laser driven lightsails, in the sail's frame: (a) For a non-rotating sail moving parallel to the beam direction, forces are balanced on both mirrors. The net force (central taupe arrow) is purely in the direction of the laser beam.  
        (b)
        Damping force: for a sail with unwanted transverse velocity component $v_y<0$, relativistic aberration angles the laser's light (green arrows), leading to non-zero transverse force opposing the transverse velocity. (c) Damping torque: For non-zero rotational velocity $\Omega^{\prime}>0$, the left  mirror  has  slightly larger  velocity away from the laser source than the right mirror, with the additional red-shift reducing the photon's momentum.   }
        \label{fig:principle}
    \end{figure}
    
Damping {\em transverse} to the direction of a light wave is also known: The Poynting-Robertson (PR) effect~\cite{Poynting1904,Robertson1937,klacka2008,klacka2014} causes dust particles orbiting a star to lose orbital angular momentum, due to relativistic aberration. In the dust's reference frame, light from the star comes from a direction shifted towards the direction the dust is moving in~\cite{EinsteinDoppler1905}. Light is absorbed, leading to a  radiation force with a component orthogonal to the radial direction to the star, slowing the dust down.
Dust particles eventually fall into the star, over a time  roughly  proportional to particle sizes and of order  tens of thousands of years~\cite{klacka2008}.
The PR effect has been shown to mildly impact solar-sail dynamics,  specifically due to residual absorption by the sail~\cite{ABDELSALAM2018897,KEZERASHVILI2013206,KEZERASHVILI20111778}. 
Can a similar effect be used for effective  damping of unwanted motion in laser-driven  lightsails? 
The situation is quite different from the usual PR effect, even as studied for solar sails: contrary to solar sails or dust, laser-powered lightsails have very small  transverse  velocities, and absorption, which is the basis of most PR studies,  must be avoided at all costs.

Here, we show that a lightsail's angular reflectivity properties, once combined with the Doppler effect and relativistic aberration, indeed provide damping similarly to the PR effect. We derive explicit expressions for the damping forces and torques for a simple two-dimensional two-mirror geometry,  and show that with appropriate optical design, transverse velocities can be damped to almost arbitrary levels by the end of the acceleration phase, albeit at the cost of increasing the acceleration distance. 

    \section{Principle}
The origins of the damping forces and torques are illustrated in Fig.~\ref{fig:principle}, using arguably the  simplest  two-dimensional reflecting object having linear mechanical stability both in translation and rotation in a laser beam: a symmetric set of two angled mirrors connected by a rod~\cite{Rafat2022}. In the figure, the desired direction (direction of the beam, $x$-axis) is upward. When moving strictly parallel to $x$, (Fig.~\ref{fig:principle}(a)), radiation pressure on both mirrors is equal and the sail is accelerated forward only. If the sail has an undesired movement orthogonal to the laser beam's direction (Fig.~\ref{fig:principle}(b)), relativistic aberration tilts the light in the sail's reference frame, so that the left mirror intercepts more light. This asymmetry leads to a component of the radiation force perpendicular to the laser beam (in the laser's frame \frameL), that is proportional and opposing the transverse velocity -- in effect a drag force.
If the sail is rotating relative to the axis of the laser beam (Fig.~\ref{fig:principle}(c)), the differential Doppler shift leads to different forces on the two mirrors, giving rise to a torque that opposes the rotational motion -- a damping torque.

\section{Notations}
    A four-vector $\vec{x}$ has, in a specified reference frame, components $(x^0,x^1,x^2,x^3)^T$ where $x^0$ is the temporal component and the other three are spatial components.  A Minkovsky metric $\text{diag}(1,-1,-1,-1)$ is implied 
    throughout. We follow the convention of using Roman indices for spatial components (\eg $x^j$) and Greek indices for all four spatio-temporal components ($x^\mu$). 
    Spatial three-vectors in any specific frame will be noted as bold Roman letters, and unit vectors will be marked with a $\hat{ } $ (\eg $\kk$ is the three-vector with components $k^i$, and $\khat=\kk/|\kk|$). Basis four-vectors of a reference frame will be denoted ${\hat{e}}_\mu$, and we will use the same notation $\mathbf{\hat{e}}_j$ for spatial unit three-vectors. We will use $\vv,v_x,v_y,v_z,v$ for the velocity vector, its components and norm respectively, in the rest frame of the laser. We will use the usual notation $\beta=v/c$, also applicable to individual components \eg $\beta_x=v_x/c$, and $\gamma=1/\sqrt{1-\beta^2}$. 
    The laser accelerating the lightsail is in frame \frameL,  assumed to be inertial, and the instantaneously co-moving inertial frame of the lightsail is called \frameM . Frame \frameM~ has velocity $\vv$ in frame \frameL. Primed quantities refer to quantities in \frameM, un-primed quantities refer to frame \frameL. The unit four-vectors in \frameM~are related to those in \frameL~  through the inverse Lorentz transform $\hat{e}'_\mu=\Lambda(-\vv)_\mu^\nu\hat{e}_\nu$, with:
    \begin{equation}
        \Lambda(\vv)=\left(
        \begin{array}{cccc}
         \gamma  & -\frac{\gamma  v_x}{c} & -\frac{\gamma  v_y}{c} & -\frac{\gamma  v_z}{c} \\
         -\frac{\gamma  v_x}{c} & 1+\frac{(\gamma -1) v_x^2}{v^2} & \frac{(\gamma -1) v_x v_y}{v^2} & \frac{(\gamma -1) v_x v_z}{v^2} \\
         -\frac{\gamma  v_y}{c} & \frac{(\gamma -1) v_x v_y}{v^2} & 1+ \frac{(\gamma -1) v_y^2}{v^2} & \frac{(\gamma -1) v_y v_z}{v^2} \\
         -\frac{\gamma  v_z}{c} & \frac{(\gamma -1) v_x v_z}{v^2} & \frac{(\gamma -1) v_y v_z}{v^2} & 1+ \frac{(\gamma -1) v_z^2}{v^2} \\
        \end{array}
        \right).\label{eq:LorentzBoost}
    \end{equation}
    
     In \frameL~we choose  $\exhat$ to point towards the desired star system.
    Note that because $\vv$ is not aligned with either $\exhat$ or $\eyhat$, the spatial parts $\exhat$ and $\exhat'$ of $\exhatf$ and $\exhatf'$ are not parallel~\cite{Frahm1998}, and neither are $\eyhat$ and $\eyhat'$.

     \begin{figure*}[btp]
        \centering
        \includegraphics{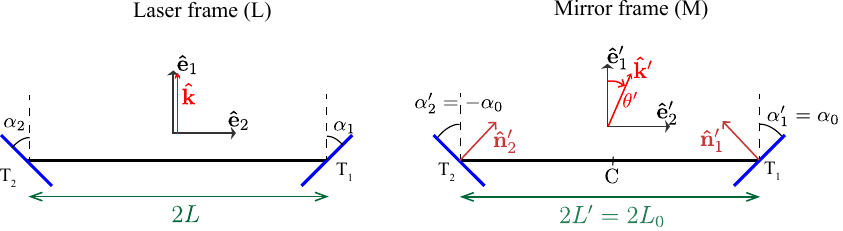}
        \caption{\label{fig:equilbirum sail}Geometry of two mirrors in the laser and mirror frames. The mirrors have surface area $A_0$ in their rest frame, C is the center of the connecting rod, and also the center of mass. The direction of $\kk$ and $\theta'>0$ on the right correspond to  $v_y<0$, that is a sail moving towards the left.}
    \end{figure*}

\section{Geometry}
     Figure~\ref{fig:equilbirum sail} shows our  lightsail model in the \frameL~and \frameM~ frames. 
     The mirrors are at the tips $T_1$ and $T_2$ of a massless rod of rest length $2L_0$, each with surface area $A_0$ (dimensions of length for our two-dimensional treatment). 
    In frame \frameM, the mirrors each make a fixed angle $\alpha_{1,2}'=\pm\alpha_0$ with $\exhat'$. In Ref~\cite{Rafat2022} this geometry was shown to have a {\em restoring} force and torque in a parabolic beam, with an equilibrium position in the middle of the beam,    and with the rod perpendicular to the beam direction.
    
    For simplicity, we  consider the laser beam to be a single plane wave with wave vector $\kk$ aligned with  $\exhat$ (Fig.~\ref{fig:equilbirum sail}), with intensity $I$ (in power per unit length, for our two-dimensional treatment). With that simplification there is no restoring force, but as we shall see there is still a damping force, and restoring and damping torques.  In \frameL, the wave four-vector  $\vec{k}$  has coordinates $k^0(1,1,0,0)^T$, with $k^0=\omega/c$.  
    
    We define  the normal vector to the mirrors, chosen to  point away from the source of light:
    \begin{equation}
     \nhat'_{1,2}(\alpha'_{1,2})=\text{sign}(\sin(\alpha'_{1,2}-\theta'))\left(\sin\alpha'_{1,2}\exhat'-\cos\alpha'_{1,2}\eyhat'\right)
    \end{equation}
    where the $\text{sign}(\sin(\alpha'_{1,2}-\theta'))$ ensures the correct orientation away from the light source.

   \section{Linear damping force}
   We aim to quantify the drag force in the $\eyhat$ direction in \frameL~that is due to the Doppler shift and relativistic aberration. To do so, we calculate the  momentum exchange rate with the sail in \frameM, which  gives us an expression of the four-force in \frameM. We then Lorentz-boost this back into \frameL, and identify the component of the force in  $\eyhat$.  As we are seeking a drag force close to the ideal trajectory, we will assume $v_y\ll v_x<c$, which, to linear order in $v_y/v$, also implies 
   \begin{gather}
   v_x/v=\sqrt{1-v_y^2/v^2}\simeq 1-\frac{1}{2}\left(\frac{v_y}{v}\right)^2 \simeq 1 \label{eq:simp_vxv}\\ 
   v_x/c \simeq \beta.  
       \label{eq:simp_betax}       
   \end{gather}
   
   \subsection{Derivation}
        In frame \frameM, in the limit of geometric optics,  forces on the mirrors can be calculated from the change of momentum of photons upon reflection. 
        In \frameM, the wave four-vector of the plane wave is given by a  Lorentz boost $k'^\mu=\Lambda({\bf v})^\mu_\nu k^\nu$ ,
        yielding
        \begin{align}
        k'^0& =Dk^0  \label{eq:k0}\\
        k'^1& =k^0\left(1+(\gamma-1)v_x^2/v^2-\gamma v_x/c\right) \simeq Dk^0 \label{eq:k1}\\
        k'^2& =k^0\left((\gamma-1)v_xv_y/v^2-\gamma v_y/c\right)  \simeq k^0\frac{v_y}{v}\left(\gamma -1-\gamma\beta\right)\label{eq:k2}
        \end{align}
        where $D$ is the Doppler factor
         \begin{equation}
        D=\gamma\left(1-\frac{\vv\cdot\khat}{c}\right)\simeq\gamma(1-\beta),
        \label{eq:dopplerfactor}
        \end{equation}
        and we have used Eqs.~(\ref{eq:simp_vxv},\ref{eq:simp_betax}) to simplify Eqs.~(\ref{eq:k1},\ref{eq:k2},\ref{eq:dopplerfactor}).
        
        The aberration angle $\theta'$ can then be obtained as
        \begin{equation}
            \theta'\simeq \tan\theta'=\frac{k'^2}{k'^1}\simeq\frac{v_y}{v}\frac{\left(\gamma -1-\gamma\beta\right)}{D}\simeq-\left(\frac{1}{D}-1\right)\frac{v_y}{v}.
            \label{eq:aberration angle}
        \end{equation}
    
         To determine the force on each mirror we multiply the momentum imparted to the mirror by each photon upon reflection by the rate of arrival of photons. In frame \frameM, each photon imparts a three-momentum 
    \begin{equation}
    \Delta\pp_M'=2\hbar(\kk'\cdot\nhat'_{1,2})\nhat'_{1,2}.
    \end{equation}
          The rate of photons  per surface area in \frameL~  is  $\Gamma=I/\hbar\omega=I/(\hbar k^0c)$, and in \frameM~becomes  $\Gamma'=D\Gamma$~\cite{McKinley1979}.  The cross-section of the mirrors intercepting the laser  is given by $A_0|\nhat'_{1,2}\cdot\khat'|$, so that in \frameM, the total three-force on each mirror is
    \begin{equation}
        \FF'_{1,2}=A_0|\khat'\cdot\nhat'_{1,2}|\Gamma'\Delta\pp_M'=2\frac{D^2I}{c} A_0|\khat'\cdot\nhat'_{1,2}|(\khat'\cdot\nhat'_{1,2})\nhat'_{1,2},\label{eq:F12}
    \end{equation}
    where we have used that, to linear order, $\kk'=Dk^0\khat'$ (Eq.~\eqref{eq:k1}). Expressing the dot product in terms of $\theta'$  and adding the contribution of both mirrors gives, to linear order in $\theta'$, the total three-force on the system:
     \begin{align}
        \FF' & \simeq 4 \frac{D^2I}{c} A_0\left(\sin^3\alpha_0 \label{eq:accelerating force}\;\exhat' + 
        \theta'\cos\alpha_0\sin 2\alpha_0 \;\eyhat'\right). 
    \end{align}    
The general expression for the coordinates of the four-force $\vec{f}$ on an object at velocity $\bf v$ in a given reference frame in terms of the three-force in that frame is $(\gamma {\bf F\cdot v}/c;\gamma {\bf F})$~\cite{faraoni_special_relativity}. In \frameM, the velocity of the sail is zero, with  $\gamma'=1$, so that  the four-force $\vec{f}$   has  coordinates  $(0;\FF')$, from which we obtain the four-force in \frameL~through Lorentz transform. Using Eq.~\eqref{eq:aberration angle} the transverse component of force (along  $\eyhat$) is then  to lowest order in $v_y/v$
\begin{align}
    f^2&\equiv \frac{dp^2}{dt'} \nonumber \\
    &\simeq  4A_0\frac{D^2I}{c}\frac{v_y}{v}\left(-\cos\alpha_0\sin 2\alpha_0(1/D-1)+
    (\gamma-1)\sin^3\alpha_0
    \right)\label{eq:Ftransverse}
\end{align}
where $t'$ is the time in \frameM, and thus proper time. We find a net transverse force that is proportional to $v_y$ and thus, depending on overall sign,  a possible drag force.

Expressing  $1/D-1$ and $\gamma -1$ for small $\beta$ gives some insight into the relative importance of the two terms in brackets in Eq.~\eqref{eq:Ftransverse}:
\begin{align}
    f^2\simeq & 4A_0\frac{D^2I}{c}\frac{v_y}{c}\left(-\cos\alpha_0\sin 2\alpha_0\left(1+\frac{\beta}{2}\right)+
    \frac{\beta}{2}\sin^3\alpha_0
    \right).
\end{align}
At small $\beta<0.2$ the second and third terms only contribute a correction of order $5\%$ to the first term. Importantly, the dominant term is negative (and independent of $v_x$), so that there is a net drag force that will lead to damping of any transverse motion, from the outset of the acceleration phase, with damping coefficient
\begin{equation}
    \zeta\equiv -f^2/v_y\simeq 4A_0\frac{D^2I}{c^2}\cos\alpha_0\sin 2\alpha_0.\label{eq:zeta}
\end{equation}
In  Eq.~\eqref{eq:Ftransverse}, the  linear $v_y/c$ dependence comes from the relativistic aberration. The $D^2I/c$ factor represents the optical power intensity in \frameM,  and in Eq.~\eqref{eq:zeta} the remaining proportionality constant $4A_0\cos\alpha_0\sin2\alpha_0$  is a geometry-dependent form of cross-section that is not simply the  radiation pressure cross-section~\cite{hulst1981light,klacka2014} but also depends on the radiation pressure's dependence on incident angle. This latter term can be adjusted by geometry, in our simple case by adjusting $\alpha_0$.

\subsection{Order of magnitude}\label{sec:transverse damping order of magnitude}
\begin{figure}
    \centering
    \includegraphics[width=0.5\textwidth]{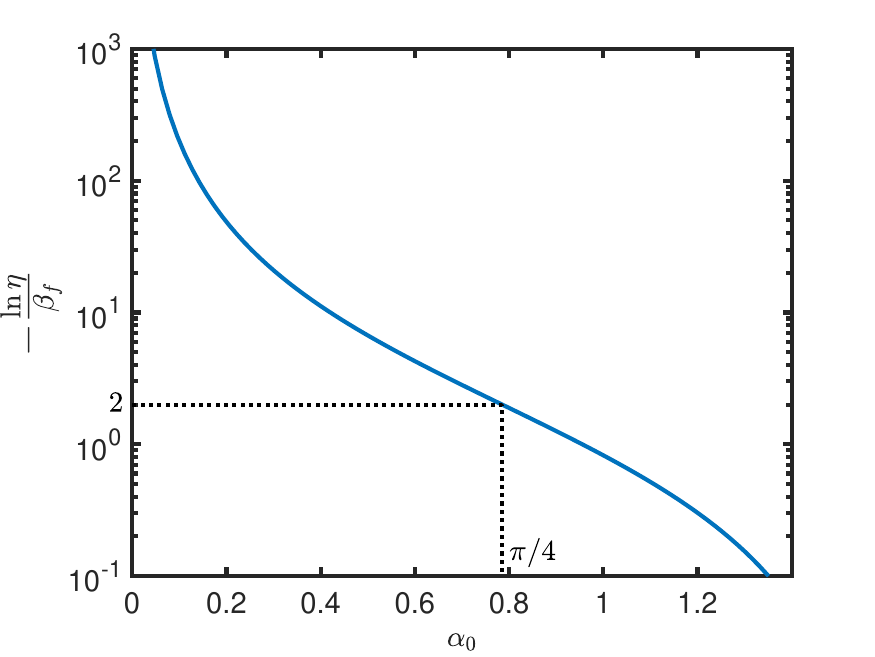}
    \caption{Ratio of transverse damping coefficient to longitudinal accelerating force as a function of mirror angle $\alpha_0$.}
    \label{fig:transverse_to_longitudinal}
\end{figure}

To assess the importance of the effect we compare the magnitude of the transverse damping force to the accelerating force along $\exhat$. 
Assuming $\beta<0.2$, we have $1\le \gamma\lesssim 1.02$ and  $0.81\simeq D\le 1$. Reasonable approximations can thus be obtained assuming $D$ constant and ignoring all other relativistic corrections
or  $\beta$ dependencies, which in particular leads to a constant longitudinal acceleration force. 
The acceleration time to reach a final $\beta_f$ can then be approximated by $t_f\simeq mc\beta_f/f^1$. 
During that time the transverse damping force leads to an exponential decrease of any initial velocity following $v_y=v_{y,0}\exp(-\zeta t/m)$.
By the end of the acceleration phase, any initial  transverse velocity $v_{y,0}$ is thus attenuated by a factor $\eta=v_{y,\rm final}/v_{y,0}$ with 
\begin{equation}
    \ln \eta=-\frac{\zeta}{m} t_f=\frac{-\zeta c}{f^1}\beta_f.
\end{equation}
From Eq.~\eqref{eq:accelerating force} and~\eqref{eq:Ftransverse} to linear order in $v_y/v$ and zeroth order in $v_x/c$ we obtain
\begin{equation}
  - \frac{\ln \eta}{\beta_f} \simeq  \frac{\cos\alpha_0\sin2\alpha_0}{\sin^3\alpha_0} =2{\rm cot}^2\alpha_0,\label{eq:transverse_to_long}
\end{equation}
which we plot in 
Fig.~\ref{fig:transverse_to_longitudinal} as a function of $\alpha_0$. For $\alpha_0=\pi/4$ and $\beta_f=0.2$ any initial velocity is thus reduced to  $\eta=\exp(-0.4)\simeq 67\%$ of its initial value. Within this approximation, valid for relatively small $\beta$, the reduction also improves exponentially with $\beta_f$.
More importantly, this reduction is geometry-dependent, with a smaller value of $\alpha_0=0.3$ leading to a final transverse velocity of only $1.5\%$ of $v_{y,0}$. However, this improvement comes at the cost of reduced longitudinal acceleration, that is, increased acceleration time and distance.

Ultimately, the damping force is a function of  the radiation pressure's angular dependence (through the aberration angle). 
Better optical design may be able to increase the level of damping without compromising acceleration time.   

\section{Restoring and damping torques}
\begin{figure}
        \centering
        \includegraphics{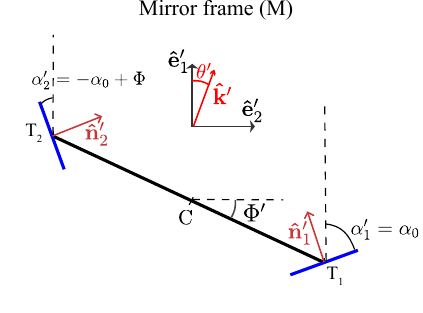}
        \caption{Rotations of the lightsail -- geometry in the mirror frame.}        \label{fig:torque_geometry}
    \end{figure}

    We now allow for rotations of the lightsail, with the axis of the sail now making an angle $\Phi'$ (Fig.~\ref{fig:torque_geometry}) with the $\bf \exhat'$ axis, with an  angular velocity $\Omega'=d\Phi'/dt'$. The vector $\vv$ now  denotes  the velocity of the center of mass C rather than the velocity of all parts of the sail.
    For $\Omega'=0$, using  Eq.~\eqref{eq:F12} to calculate the torque about C in \frameM, we get,  to linear order in $\theta'$ and $\Phi'$ :
    \begin{equation}
        \tau'_r=-\left(\frac{4L_0A_0ID^2}{c}\sin2\alpha_0 \right)\left(\Phi'-\theta'\right){\bf \ezhat'},\label{eq:restoring torque}
    \end{equation}
    which is a restoring torque towards the direction of the incoming light in \frameM. To obtain a damping term for $\Omega'\neq 0$ we need to take into account the rotational velocities of the mirrors, which lead to different Doppler shifts and relativistic aberrations for each point of each  mirror.   
    We take the further simplifying assumption that the mirrors  are  small compared to $L_0$, and can thus be treated as point mirrors, each with unique velocities $\vv'_{1,2}$ in \frameM.  Equation~\eqref{eq:F12} remains valid, but we now have two different values $\khat'_{1,2}$ and $D_{1,2}$. Since $v_x$  can reach relativistic values, we need to use relativistic velocity additions $\vv_{1,2}=\vv\oplus \vv'_{1,2}$. Assuming $v'_{1,2}\ll c$, we obtain to linear order 
     \begin{equation}
              \vv_{1,2}\simeq \vv + \frac{v'_{1,2x}}{\gamma^2}{\bf \exhat}+\frac{v'_{1,2y}}{\gamma}{\bf \eyhat},\ \label{eq:velocity composition lowest order final}
    \end{equation}
    from which, again to linear order in $v'_{1,2}/c$ and $v_y/v$, we obtain
    \begin{equation}
    D_{1,2}  \simeq D\left(1-
                            \frac{v'_{1,2x}}{c}\right)
    \end{equation}
    and 
     \begin{equation}
        \theta'_{1,2}\simeq -\left(\frac{1}{D}-1\right)\left(\frac{v_y}{v}+\frac{v'_{1,2y}}{\gamma v}\right). \label{eq:aberration angle tips lowest order}
    \end{equation}
    The damping torque at ``equilibrium'' is calculated for $v_y=0$, $\Phi'=0$. In \frameM, the velocities of the mirrors then simplify to  $\vv'_{1,2}= \Omega'{\bf \ezhat'}\times{\bf CT}_{1,2}=\mp\Omega'L_0{\bf \exhat}'$, so that $\theta'_{1,2}=\theta'$ and $D_{1,2}=D(1\pm\Omega'L/c)$. The total  torque becomes, again to linear order,
    \begin{equation}
        {\bf \tau}'_d=-\left(\frac{8L_0^2A_0ID^2}{c^2}\sin^3\alpha_0\right) \Omega'{\bf \ezhat'} \label{eq:damping torque},
    \end{equation}
    and is indeed a damping torque as it is countering and proportional to $\Omega'$. 

    Equations~\eqref{eq:restoring torque} and \eqref{eq:damping torque} lead to a damped oscillating rotational motion, with damping time constant $t'_{rd}$. Using the total moment of inertia $mL_0^2$ we have 
    \begin{equation}
    t'_{rd}  =\frac{mc^2}{2A_0ID^2\sin^3\alpha_0}  
    \end{equation}
which, by a reasoning analogous to that in Section~\ref{sec:transverse damping order of magnitude} leads to a rotational damping ratio $\eta_r$ with 
\begin{equation}
\ln \eta_r=-\beta_f/2.\label{eq:eta_r}
\end{equation}
Eq.~\eqref{eq:eta_r} does not depend on $\alpha_0$. However the damping ratio depends on the ratio of longitudinal forces to  moment of inertia, so that it does in general, depend on geometry. For the two-mirror system under consideration with a final velocity of $\beta_f=0.2$, any initial rotational velocity would be reduced by only 10\%, but this could be enhanced considerably by having a mass distribution closer to the center of gravity ({\em e.g}. a payload at the center), or via different optical designs.

\section{Conclusion}
The relativistic transformation of photon momentum can  contribute to significant damping of unwanted  transverse and rotational movements of lightsails.
Both the space-like and time-like transformations are of importance: For our geometry around equilibrium, transverse {\em translational} damping comes primarily from the relativistic aberration (space-like) while {\em rotational} damping comes primarily from  the difference in Doppler terms for both mirrors (time-like). However, for different values of $\Phi'$ (or  different geometries), both the relativistic aberration and Doppler effects can be at play for the translational and rotational damping, with similar magnitude. 
Order of magnitude calculations show  that over the course of the acceleration phase, unwanted transverse velocity components can be damped to almost arbitrary levels by appropriate optical design (Fig.~\ref{fig:transverse_to_longitudinal}). In our simple geometry,  this comes at the cost of slower acceleration, but as transverse damping ultimately comes from the angular response of the reflection, it is conceivable that very large damping coefficients could be achieved without loss of longitudinal acceleration in  realistic geometries, for example using  gratings or meta-surfaces with sharp angular diffraction patterns~\cite{Ilic2019, Salary2020,Kumar2021}.

Our study also highlights that taking the full relativistic transformations of light into account is essential to include damping mechanisms in lightsail dynamics.
Studies of lightsails we have encountered so far have included the time-like (Doppler) transformation, but neglected the space-like (relativistic aberration) transformation.

Finally, among our many simplifying assumptions,  we considered  the laser to be a  single monochromatic plane-wave. In practice, the divergence of laser beams will be comparable to the magnitude of the relativistic aberration. 
When extending this  study  to more realistic lightsail geometries, it will thus be important to also take into account the angular spread of finite beams, noting the relativistic transformation of finite beams is a rich field of study in itself, with many surprising phenomena~\cite{McKinley1979,Yessenov2023}.  

That the Doppler effect and relativistic aberration can in principle provide damping should be welcomed by the interstellar lightsail community, as other mechanisms of damping residual motion seem difficult to implement within the extremely narrow mass budget constraint. Special relativity gifts us a solution that could simplify lightsail designs and considerably improve accuracy of trajectories.
\begin{acknowledgments}
We thank Mr Ronan Potts for help with early exploratory work,  Profs Peter Tuthill and Martijn de Sterke for useful discussions.
\end{acknowledgments}

\bibliography{dopplerAPL}
\end{document}